\documentclass[prl,amsfonts,showpacs,showkeys,preprint]{revtex4}
\usepackage{graphicx}
\usepackage{dcolumn}
\usepackage{bm}

\RequirePackage{times}
\RequirePackage{mathptm}

\begin{document}

\title{On the solution of trivalent decision problems by quantum state identification}

\author{Karl Svozil}
\email{svozil@tuwien.ac.at}
\homepage{http://tph.tuwien.ac.at/~svozil}
\affiliation{Institut f\"ur Theoretische Physik, University of Technology Vienna,
Wiedner Hauptstra\ss e 8-10/136, A-1040 Vienna, Austria}
\author{Josef Tkadlec}
\email{tkadlec@fel.cvut.cz}
\homepage{http://math.feld.cvut.cz/tkadlec/}
\affiliation{Department of Mathematics, Faculty of Electrical Engineering,
Czech Technical University, 166\,27 Praha, Czech Republic}

\begin{abstract}
The trivalent functions of a trit can be grouped into equipartitions of three elements. We discuss the separation of the corresponding functional classes by quantum state identifications.
\end{abstract}

\pacs{03.67.Lx,03.65.Ud}
\keywords{Trits, trivalent decision problems, quantum state identification}

\maketitle

One of the advantages of quantum computation \cite{nielsen-book,Gruska,benn:97,Ozhigov:1997,bbcmw-01,cleve-99,fortnov-03}
over classical algorithms \cite{rogers1,odi:89} is due to
the fact that in quantum mechanics information
can be coded in or ``spread among'' coherent states in such a way that
certain decision problems can be solved by identifying a quantum state
which ``globally'' contains the solution \cite{mermin-02,svozil-2005-ko}.
Thereby,
information about single cases are not useful for (and even makes impossible)
a decryption of the quantum computation.
This feature is not only present in binary decision problems of the usual type,
such as Deutsch's algorithm, but can be extended to d-ary decision problems on dits.

In what follows we shall consider as the simplest of such problems
the trivalent functions of trits.
We shall group them in three functional classes
corresponding to an equipartition of the set of functions into three elements.
We then investigate
the possibility to separate each of these classes by
quantum state identifications \cite{DonSvo01,svozil-2002-statepart-prl}.

Formally, we shall consider the functions $f {\,{:}\linebreak[0]\,\;}\{-,0,+\} \to \{-,0,+\}$.
There are $3^3 =
27$ such functions.
The dits will be coded by elements of some  orthogonal base in ${\mathbb{C}}^3$.
Without loss of
generality we may take $(1,0,0) = |-\rangle$, $(0,1,0) = |0\rangle$, $(0,0,1) = |+\rangle$.
We will be searching for a function $g {\,{:}\,\;} \{-,0,+\} \to {\mathbb{C}}$ such that the
images of the mapping $\{-,0,+\}^{\{-,0,+\}} \to {\mathbb{C}}^3$ defined by
  $$f \mapsto g(-)\,|-\rangle + g(0)\,|0\rangle + g(+)\,|+\rangle$$
form the smallest possible number of orthogonal triples.

First, let us show that we may find a function $g$ such that we obtain 3
orthogonal triples. Let the values of $g$ be the $\sqrt[3]{1}$ (in the
set of complex numbers) and put, e.g.,
  $$g(x) = {\rm e}^{2\pi{\rm i}x/3}$$
Let us, for the sake of simplicity and brief notation,
identify `$-$' with `$-1$' and `$+$' with `$+1$,'
and
denote $\alpha = {\rm e}^{2\pi {\rm i}/3}$ and $\overline\alpha = \alpha^2$. Hence, $\alpha^3=1$,
$\alpha {\overline\alpha}= 1$, $\alpha + \overline\alpha= -1$.
Then, the ``quantum oracle'' function $g$ is given
by the following table:
  $$
  \begin {array}{c||c|c|c}
  x    & -       & 0 & +\\\hline
  g(x) & \overline\alpha & 1 & \alpha
  \end {array}
  $$

The following triples of functions can be assigned the same vector (except a nonzero multiple)
by the following scheme:
  $$
  \begin {array}{|*3{@{\;(}c@{,}c@{,}c@{)\;\;}c@{\;\;}c|}}
  \hline
  -&-&- &       &                    & -&-&0 &       &              & -&-&+ &       & \\
  0&0&0 &\mapsto& (1,1,1)            & 0&0&+ &\mapsto& (1,1,\alpha) & 0&0&- &\mapsto& (1,1,\overline\alpha)\\
  +&+&+ &       &                    & +&+&- &       &              & +&+&0 &       & \\
  \hline
  -&0&+ &       &                    & -&0&- &       &              & -&+&- &       & \\
  0&+&- &\mapsto& (1,\alpha,\overline\alpha) & 0&+&0 &\mapsto& (1,\alpha,1) & 0&-&0 &\mapsto& (1,\overline\alpha,1)\\
  +&-&0 &       &                    & +&-&+ &       &              & +&0&+ &       & \\
  \hline
  -&+&0 &       &                    & 0&-&- &       &              & +&-&- &       & \\
  +&0&- &\mapsto& (1,\overline\alpha,\alpha) & +&0&0 &\mapsto& (\alpha,1,1) & -&0&0 &\mapsto& (\overline\alpha,1,1)\\
  0&-&+ &       &                    & -&+&+ &       &              & 0&+&+ &       & \\
  \hline
  \end {array}
  $$
In every column we obtain an orthogonal triple of vectors. Moreover,
vectors from different orthogonal triples are apart by the same angle $\phi$, for which
$\cos \phi = \sqrt3/3$.

Now, let us prove by contradiction that the function $g$ cannot be defined in such a way that
we obtain at most two orthogonal triples of subspaces.
For the sake of contradiction, let us suppose that this proposition is false.

First, all values $g(-), g(0), g(+)$ should be nonzero (if, e.g., $g(-) = 0$
then the vector $\bigl(g(-),g(-),g(-)\bigr)$ assigned to the function
$(-,-,-)$ is a zero vector). Hence, we obtain a linear subspace generated by
the vector $(1,1,1)$.

Second, $g(-), g(0), g(+)$ cannot have the same value (in this case we
obtain only one subspace generated by the vector $(1,1,1)$).

Let us show that the vectors assigned to the functions $(-,-,0)$ and
$(-,0,0)$ are not orthogonal. Indeed, if they are orthogonal, then $0 = g(-)
\, \overline {g(-)} + g(-) \, \overline {g(0)} + g(0) \, \overline {g(0)} =
|g(-)|^2 + g(-) \, \overline {g(0)} + |g(0)|^2$ and therefore $g(-) \,
\overline {g(0)}$ is a negative real number. Hence $0 = |g(-)|^2 - |g(-)|
\cdot |g(0)| + |g(0)|^2 = \bigl( |g(-)| - \frac 12 \, |g(0)| \bigr)^2 +
\frac 34 \, |g(0)|^2$ and therefore $g(0) = 0$ that is impossible.

Let us show that all values $g(-), g(0), g(+)$ are different. Indeed, let,
e.g., $g(-) = g(0)$. Analogously as in the previous paragraph we can show
that the vectors $\bigl( g(-), g(-), g(+) \bigr)$ and $\bigl( g(-), g(+),
g(+) \bigr)$ are not orthogonal. These vectors do not generate the same
subspace (otherwise $g(-) = g(0) = g(+)$) and are not multiples of the
vector $(1,1,1)$, hence at least one of them should be orthogonal to
$(1,1,1)$. Let, e.g., $\bigl( g(-), g(-), g(+) \bigr)$ is orthogonal to
$(1,1,1)$. Then $2\,g(-) + g(+) = 0$ and therefore this vector is a multiple
of $(1,1,-2)$. The subspace making an orthogonal triple vith subspaces
generated by vectors $(1,1,1)$ and $(1,1,-2)$ is generated by $(1,-1,0)$.
But this is impossible because no coordinate can be zero.

We have shown that the subspaces assigned to functions $(-,-,0)$ and
$(-,0,0)$ are not orthogonal and do not coincide (otherwise $g(-) = g(0)$).
Hence they do not belong to one orthogonal triple and at least one of them
should belong to an orthogonal triple with the space generated by the
vector $(1,1,1)$. Let, e.g., $\bigl( g(-), g(-), g(0) \bigr)$ is orthogonal
to the vector $(1,1,1)$. Then $2 \, g(-) + g(0) = 0$. Analogously to
previous paragraphs we can show that one of the vectors $\bigl( g(-), g(-),
g(+)\bigr)$ and $\bigl( g(-), g(+), g(+) \bigr)$ ($\bigl( g(0), g(0), g(+)
\bigr)$ and $\bigl( g(0), g(+), g(+) \bigr)$, resp.) is orthogonal to the
vector $(1,1,1)$. Since the values of the function $g$ are different, we
obtain $g(-) + 2 \, g(+) = 0$ and $2\,g(0) + g(+) = 0$. The system of
equations has the only solution $g(-) = g(0) = g(+) = 0$,
which results in a complete contradiction.

In summary we find that we cannot solve the type of trivalent decision problems
as discussed above by a single query.
Such a behavior has already been observed for the problem to find the
parity of an unknown binary function $f:\{0,1\}^k \rightarrow \{0,1\}$ of $k$ bits,
which turned out to be quantum computationally hard
\cite{Farhi-98,bbcmw-01,Miao-2001,orus-04,stadelhofer-05}.
We conjecture that this hardness increases with the number $d$ of possible states of a single bit.

\section *{Acknowledgements}

The work was supported by the research plan of the Ministry of Education of
the Czech Republic no.~6840770010 and by the grant of the Grant Agency of
the Czech republic no.~201/07/1051 and by a the exchange agreement
of both of our universities.


\end {document}